\documentclass[aps,prl,twocolumn,floatfix,groupedaddress,superscriptaddress,footinbib,notitlepage,showpacs]{revtex4-1}
\usepackage{graphicx,graphics,times,bm,bbm,bbold,amssymb,amsmath,amsfonts,dsfont,color,xcolor,dcolumn}
\usepackage[bookmarksnumbered, colorlinks,plainpages]{hyperref}
\hypersetup{colorlinks,linkcolor=blue,citecolor=blue,urlcolor=blue}
\newtheorem{theorem}{Theorem}

\newcommand{\ket}[1]{|#1\rangle}
\newcommand{\bra}[1]{\langle #1|}
\newcommand{\project}[1]{\ket{#1}\bra{#1}}

\newcommand{\Tr}{{\mathrm {Tr}}}
\newcommand{\diag}{{\mathrm {diag}}}

\let\oldsqrt\sqrt
\def\sqrt{\mathpalette\DHLhksqrt}
\def\DHLhksqrt#1#2{%
\setbox0=\hbox{$#1\oldsqrt{#2\,}$}\dimen0=\ht0
\advance\dimen0-0.2\ht0
\setbox2=\hbox{\vrule height\ht0 depth -\dimen0}%
{\box0\lower0.4pt\box2}}

\DeclareFontFamily{OT1}{pzc}{}
\DeclareFontShape{OT1}{pzc}{m}{it}%
              {<-> s * [1.25] pzcmi7t}{}
\DeclareMathAlphabet{\mathpzc}{OT1}{pzc}%
                                 {m}{it}

\begin{document}

\title{How Quantum is a ``Quantum Walk"?}

\author{F. Shahbeigi}
\affiliation{Department of Physics, Ferdowsi University of Mashhad, Mashhad, Iran}

\author{S. J. Akhtarshenas}
\email{akhtarshenas@um.ac.ir}
\affiliation{Department of Physics, Ferdowsi University of Mashhad, Mashhad, Iran}

\author{A. T. Rezakhani}
\affiliation{Department of Physics, Sharif University of Technology, Tehran 14588, Iran}

\begin{abstract}
We characterize quantumness of the so-called quantum walks (whose dynamics is governed by quantum mechanics) by introducing two computable measures which are stronger than the variance of the walker's position probability distribution. The first measure is based on comparing probability distributions of a quantum walk and all classical random walks (through the classical relative entropy of the distributions), and it quantifies non-Gaussianity of the probability distribution of the walk. Next, after assigning a density matrix to classical walks, we introduce a more powerful measure by employing quantum relative entropy. We show that this measure exceeds the first one by the quantum coherence of the walk. There are walks labeled classical by the variance whereas our measures identify some quantumness therein. As an application, we study a model of quantum (energy) transport on a simple lattice, and compare the behavior of its relative transport efficiencies with that of the quantumness. Our measures help partly explain why in some quantum transport phenomena a considerably high efficiency may appear---this is where quantumness is appreciable.
\end{abstract}

\maketitle

\textit{Introduction.---}Random walk (RW), one of the most fundamental subjects in probability theory, plays a key role in diverse fields of science (mathematics, physics, chemistry, biology, medicine, economics, and social sciences) \cite{rw-book, book:reichl, chandrasekhar, chem, chem-2, wright, rw-biology:book, nosofsky, rw-economics}. It is usually employed to describe phenomena which exhibit stochasticity in their behavior. In its most elementary version, RW includes a walker moving on a line (or a one-dimensional curve) in a discrete manner, changing its position $\textit{x}$ (initially at $\textit{x}_{0}=0$) one step to the right or left depending on the result of flipping a coin at each step; e.g., heads for right and tails for left. This associates a probability distribution (say $\textit{p}_{+}$ and $\textit{p}_{-}=1-\textit{p}_{+}$) to the moves, which in turn yields a probability distribution for $\textit{x}$ after $\tau$ time steps, $\textit{P}_{\textsc{rw}}(\textit{x},\tau;0,0)= \binom{\tau}{(\tau+\textit{x})/2}\textit{p}_{+}^{(\tau+\textit{x})/2} \textit{p}_{-}^{(\tau-\textit{x})/2}$. This model has been generalized in numerous respects, e.g., in several dimensions, on arbitrary graphs or surfaces, in continuous space-time, or with sophisticated dynamics \cite{rw-fractal, rw-fb,rw-complex}.

A particularly interesting and powerful generalization of RW is quantum walk (QW) which is obtained by imposing quantum mechanics on the walker's dynamics \cite{AharonovPRA1993,farhi contin}. QW has garnered numerous applications in understanding underlying principles in various problems in physics, quantum computation,  and quantum information theory \cite{d-aharonov,ambainis-1d, aaronson, childs pra2004 1, shenvi pra2003,farhi thcom2008, ambainis2003,childs-qip-qalgorithm,childs2009prl,childs-science,omar-1, omar-2, qw-book,logic-walk,walk-bec,science-strong, sciarrino}. For example, QW is useful development of novel and fast quantum algorithms \cite{childs-qip-qalgorithm,aaronson, childs pra2004 1, shenvi pra2003,qw-book}, explanation of (quantum) transport phenomena in complex structures \cite{mulken}, or energy transfer in photosynthetic light-harvesting complexes \cite{dorner, pleniosuasana, chandrashekar2, chin, mohseni, Panitchayangkoona}. 

It is a general consensus in quantum information science that quantum protocols may offer substantial enhancement compared to their classical counterparts \cite{book:Nielsen}. Nevertheless, and despite vast efforts, it is yet not decisively concluded what underlie(s) such relative power and strength \cite{strength,preskill}. A natural step toward answering this question seems to be development of methods and tools to discern genuinely quantum mechanical from classical features. In the context of QW, it has been observed that RW and QW are different in their probability distributions ($\textit{P}_{\textsc{rw}}(\textit{x},\tau)$ and $\textit{P}_{\textsc{qw}}(\textit{x},\tau)$, where we have lightened the notation and omitted the initial condition $(\textit{x}_{0},\tau_{0})=(0,0)$). As a result, the $\tau$-dependence of the variance of the walker's position $\sigma^{2}(\tau)=\langle \textit{x}^{2}(\tau) \rangle - \langle \textit{x}(\tau) \rangle^{2}$ can be different; $\sigma^{2}_{\textsc{rw}}(\tau)\propto \tau$ whereas $\sigma^{2}_{\textsc{qw}}(\tau)$ in general shows a more complex behavior which in asymptotically large $\tau$s (long-time limit) becomes quadratic ($\sigma^{2}_{\textsc{qw}}(\tau)\propto \tau^{2}$, for $\tau\gg1$) \cite{meyer}. This sharp difference in $\sigma^{2}(\tau)$ for RW and QW has motivated its use as an indicator to discern these two walk models \cite{brunprl2003, shapira, romanelli 2, rezakhani}. Despite this stark difference, there is evidence that one cannot always take $\sigma^{2}$ as faithful measure to distinguish RW and QW. It seems that there is no well-founded reason to prove that a process whose variance is linear in time is definitely obtained by a genuinely ``classical" process in the full absence of any quantum effect. From a straightforward calculation one can see that $\sigma^{2}_{\textsc{rw}}(\tau)=4\tau\, \textit{p}_{+}\textit{p}_{-}$, which has the maximum slope $1$, thus one may infer that at least the processes whose variances are linear in $\tau$ with the slope $>1$ are not evidently classical. Note that, relying on $\sigma^{2}(\tau)$ all QWs in the presence of decoherence on the coin are characterized as RWs \cite{brunprl2003}. Nevertheless, interaction with an environment does not necessarily kill all (useful) quantum effects or resources; in some cases the environment may even come to assist \cite{mohseni,Amin,chin,Kasal,ekert-papers}.

In this paper, we introduce two \textit{computable} measures which identify what possibly gives the QW model its strengths---``quantumness." Our first quantumness measure $\textit{Q}(\tau)$ is defined as the minimum relative entropy of the probability distributions $\textit{P}_{\textsc{qw}}$ and $\textit{P}_{\textsc{rw}}$ where the minimum is taken over all RWs. A complete description of QW requires the use of density matrices rather than the probability distribution formulation. We first associate a physically justifiable density matrix to an RW, whose diagonal entries are given by $\textit{P}_{\textsc{rw}}(\textit{x},\tau)$ of the RW. In fact, we show that classical RW density operator is obtained by the action of a quantum channel which kills all coherence in a density matrix. We then define our second (stronger) genuine quantumness measure $\mathpzc{Q}(\tau)$ as the minimum quantum relative entropy of our QW with respect to the set of all classical RWs. Intuitively, $\mathpzc{Q}(\tau)$ contains more quantumness than $\textit{Q}(\tau)$; we prove that this excessive value is exactly equal to the quantum coherence of the quantum walker measured by the relative entropy of coherence. Although our focus here is on discrete-time QW, extension to continuous-time QW seems also straightforward especially due to the correspondence of discrete-time and continuous-time QWs \cite{childs-cmp}.


\textit{Quantum walk.---}Suppose that $\{|\textit{x}\rangle, \textit{x}\in\mathbb Z\}$ is the eigenstates of the walker's position operator, yielding an orthonormal basis for the walker's Hilbert space $\mathpzc{H}_{\,\mathrm{w}}$. The walk is driven by a coin, a two-dimensional quantum system spanned by the orthonormal basis $\{|\pm\rangle\}$ (defining $\mathpzc{H}_{\,\mathrm{c}}$). Each step of the QW is given by the unitary
\begin{equation}
\label{qw evo ope}
\textit{W}= \textit{E}(\textit{U}\otimes\mathbbmss{I}),
\end{equation}
acting on $\mathpzc{H}_{\,\mathrm{c}} \otimes \mathpzc{H}_{\,\mathrm{w}}$, where $E= \sum_{i=\pm}|i\rangle\langle i|\otimes \textit{M}_{i}$, $\textit{M}_{\pm}$ (where $\textit{M}_{-}=\textit{M}^{\dag}_{+}$) are the right- and left-move operators, $\textit{M}_{\pm} |\textit{x}\rangle=|\textit{x}\pm1\rangle$, and $U$ is the unitary operator of the coin.

Starting from a pure coin-walker state $\ket{\Psi(0)}$, the state of the combined system evolves as $\ket{\Psi(\tau)}=\textit{W}^{\tau} \ket{\Psi(0)}$ \cite{brunprl2003, montero pra,Nayak2000,ambainis-1d}. This state can be represented as
\begin{equation}
\label{CoinParticlePsi}
\ket{\Psi(\tau)}=\ket{+}\ket{\Psi_{+}(\tau)}+\ket{-}\ket{\Psi_{-}(\tau)},
\end{equation}
where $\ket{\Psi_{+}(\tau)}$ and $\ket{\Psi_{-}(\tau)}$ are two unnormalized and nonorthogonal states in $\mathpzc{H}_{\,\mathrm{w}}$ whose analytic expressions (in terms of hypergeometric functions) we give in Ref. \cite{SupplementalMaterial}. Given the coin-walker state \eqref{CoinParticlePsi}, we can obtain the walker's density matrix by tracing out over the coin, 
\begin{equation}
\label{ParticleState}
\varrho_{\textsc{qw}}{(\tau)}=\ket{\Psi_{+}(\tau)}\bra{\Psi_{+}(\tau)}+\ket{\Psi_{-}(\tau)}\bra{\Psi_{-}(\tau)}.
\end{equation}
More generally, environmental effects at each step of the walk can be realized by some noise operation acting on the coin-walker system before the operator $W$ \cite{brunprl2003,romanelli 2, lopez, annabestani}. For example, consider noise acts only on the coin, where the evolution of the QW is modified to be given by the following operation (rather than simply by the unitary map $\mathpzc{W}(\cdot)=\textit{W}\cdot \textit{W}^{\dag}$): 
\begin{equation}\label{QW and env evo }
\mathpzc{W}_{\mathrm{noisy}}= \mathpzc{E}(\mathpzc{U}\,\mathpzc{N}\otimes\mathpzc{I}),
\end{equation}
where $\mathpzc{E}(\cdot)=\textit{E}\cdot \textit{E}^{\dag}$, $\mathpzc{U}(\cdot)=\textit{U} \cdot \textit{U}^{\dag}$, $\mathpzc{N}$ is a quantum noise channel on $\mathpzc{H}_{\,\mathrm{c}}$, and $\mathpzc{I}$ is the identity channel on $\mathpzc{H}_{\,\mathrm{w}}$.


\textit{Quantumness of QWs.---}We introduce a measure of quantumness of a QW based on the relative entropy of the walker's probability distributions. Relative entropy has been widely used in quantifying various quantum notions such as quantum entanglement \cite{vedral97}, quantum discord \cite{modi}, and quantum coherence \cite{pleniocoherence2014}. For two probability distributions $\textit{P}=\{\textit{p}_i\}_{i}$ and $\textit{P}'=\{\textit{p}'_{i}\}_{i}$, the relative entropy is defined by $\textit{D}(\textit{P}\Vert \textit{P}')=\sum_i \textit{p}_i(\log \textit{p}_i-\log \textit{p}'_i)$ (recall that $\textit{H}(\textit{P})=-\sum_{i} \textit{p}_i\log \textit{p}_i$ is the Shannon entropy \cite{note}) and measures how different they are. For a given QW with associated position probability distribution $\textit{P}_{\textsc{rw}}(\tau)=\{\textit{P}_{\textsc{rw}}(\textit{x},\tau)\}_{\textit{x}}$, one can define a measure of quantumness as
\begin{align}
\label{p rho}
\textit{Q}(\tau)=\min_{\textit{P}_{\textsc{rw}}(\tau)} \textit{D}\big(\textit{P}_{\textsc{qw}}(\tau)\|\textit{P}_{\textsc{rw}}(\tau)\big),
\end{align}
where the minimization is taken over position probability of all classical RWs, or equivalently, over all possible probabilities $\textit{p}_{+}$ for the coin in an RW process. This optimization can be performed in a straightforward manner for the case the walker is initially localized at $\textit{x}=0$---see the discussion at the end of the paper for more general cases. From $\textit{P}_{\textsc{rw}}(\textit{x},\tau)$  
we can see that \cite{SupplementalMaterial} the minimum is attained if the RW is driven by a coin with probability $\textit{p}_{+}^{*}(\tau)=1/2+\langle \textit{x}(\tau)\rangle_{\textsc{qw}}/(2\tau)$. 
For this coin, the first moment of the RW is given by
\begin{align}
\label{Reference}
\langle \textit{x}(\tau)\rangle^{*}_{\textsc{rw}}
=\langle \textit{x}(\tau)\rangle_{\textsc{qw}}.
\end{align}
The following theorem provides a closed form for $\textit{Q}(\tau)$.

\begin{theorem}
\label{LemmaRef}
For a QW described by the position distribution $\textit{P}_{\textsc{qw}}(\tau)$, the quantumness $\textit{Q}(\tau)$ is given by
\begin{align}
\textit{Q}(\tau)&= \textit{D}\big(\textit{P}_{\textsc{qw}}(\tau)\|\textit{P}_{\textsc{rw}}^{*}(\tau)\big)  \nonumber\\
&= \textit{D}\big(\textit{P}_{\textsc{qw}}(\tau)\|\textit{P}_{\textsc{rw}}^{\textsc{1/2}}(\tau)\big)-\textit{D}\big(\textit{P}_{\textsc{rw}}^\ast(\tau)\|\textit{P}_{\textsc{rw}}^{\textsc{1/2}}(\tau)\big),
\end{align}
    where $\textit{P}_{\textsc{rw}}^{\ast}(\tau)$ and $\textit{P}_{\textsc{rw}}^{\textsc{1/2}}(\tau)$ are the position distribution of two RWs with associated coin probabilities $\textit{p}_{+}=\textit{p}_{+}^{*}$ and $\textit{p}_{+}=1/2$, respectively.
\end{theorem}
For a proof see Ref. \cite{SupplementalMaterial}. An immediate consequence is that
\begin{equation}
\label{upperbound}
\textit{Q}(\tau)\leqslant \textit{D}\big(\textit{P}_{\textsc{qw}}(\tau)\| \textit{P}_{\textsc{rw}}^{\textsc{1/2}}(\tau)\big).
\end{equation}
Recalling Eq. (\ref{Reference}), the upper bound is achieved if $\langle \textit{x}(\tau)\rangle_{\textsc{qw}}=0$. For example, the Hadamard coin ($\textit{U}=(1/\sqrt{2})\left(\begin{smallmatrix}1& 1\\ 1& -1\end{smallmatrix} \right)$ in the $\{|\pm\rangle\}$ basis) with the initial state $\ket{\phi_0}=(\ket +\pm i\ket -)/\sqrt{2}$ and the identity coin ($\textit{U}=\mathbbmss{I}$) with the initial state $\ket{\phi_0}=(\ket + +\textit{e}^{i\gamma}\ket -)/\sqrt{2}$ saturate the upper bound (for any $\gamma$).

Interestingly, Eq. \eqref{Reference} shows that the nearest RW reference to a given QW is the one that possesses the same first moment as the one of the QW. Recalling that in the long-time limit ($\tau\gg1/\textit{p}_{+}$) the RW distribution becomes Gaussian ($\textit{P}_{\textsc{rw}}(\textit{x})\propto \textit{e}^{-(\textit{x}-\langle \textit{x}\rangle)^{2}/(2\sigma^{2})}$), 
the quantumness $\textit{Q}(\tau)$ of a QW in some sense can be considered as a measure of non-Gaussianity of the QW's probability distribution. However, unlike usual cases where the Gaussian reference of a probability distribution is expected to have the same first and second moments as the original distribution, here the Gaussian reference distribution of a QW (denoted by $\textit{P}_{\textsc{g}}^\ast(\tau)$) coincides with the QW's distribution in the first moment (note that for this case the second moment of an RW depends on the first moment; $\langle \textit{x}^2(\tau)\rangle_{\textsc{rw}}=\tau(1-\langle \textit{x}(\tau)\rangle^{2}_{\textsc{rw}}/\tau^{2}) + \langle \textit{x}(\tau)\rangle^{2}_{\textsc{rw}}$). More precisely, for a unitary QW process with the coin initially in
\begin{equation}
\label{phi0}
\ket{\phi_0}=\cos\eta \ket{+}+ \textit{e}^{i\gamma}\sin\eta \ket{-},
\end{equation}
and updated by the unitary operation
\begin{align}
\label{Uc}
\textit{U}=\begin{pmatrix}
\textit{e}^{i\alpha}\cos\theta & \textit{e}^{-i\beta}\sin\theta\\
\textit{e}^{i\beta}\sin\theta & -\textit{e}^{-i\alpha}\cos\theta\\
\end{pmatrix},\,\,0\leqslant\theta\leqslant\pi/2,
\end{align}
the Gaussian reference is determined by a coin described by the probability $\textit{p}^{*}_{+}= 1/2+(\cos2\eta + \tan\theta\cos\varphi \sin 2\eta)(1 - \sin\theta)/2$, where $\varphi=\alpha+ \beta-\gamma$ (obtained by using results of Refs. \cite{brunprl2003,montero pra}).

Furthermore, in the long-time limit, Eq. \eqref{p rho} yields
\begin{align}
\label{large-time}
\textit{Q}(\tau)
&\approx \textit{D}\big(\textit{P}_{\textsc{qw}}(\tau)\|\textit{P}_{\textsc{g}}^\ast(\tau)\big)\\
&= \textit{H}\big(\textit{P}_{\textsc{g}}^\ast(\tau)\big)-\textit{H}\big(\textit{P}_{\textsc{qw}}(\tau)\big)+\frac{\sigma_{\textsc{qw}}^2(\tau)-\sigma_{\textsc{g}}^{\ast2}(\tau)}{2\sigma_{\textsc{g}}^{\ast2}(\tau)}, \nonumber
\end{align}
where $\sigma_{\textsc{g}}^{\ast2}(\tau)$ is the position variance of the Gaussian reference distribution. In the case that $\sigma_{\textsc{qw}}^2(\tau)=O(\tau^{2})$, Eq. \eqref{large-time} shows that the asymptotic behavior of $\textit{Q}(\tau)$ is $O(\tau)$ because $\sigma^{2}_{\textsc{rw}}(\tau)=O(\tau)$. Moreover, even if $\sigma_{\textsc{qw}}^2(\tau)=O(\tau)$, nonvanishing quantumness may still survives.


\textit{Total quantumness.---}Thus far, we have provided a measure which quantifies how different a QW is from RWs based on comparing the position probability distribution of a QW walker with that of RWs. This picture may, however, be incomplete. In fact, although for classical RWs, the walker's position distribution provides a complete description of the process, it is not necessarily the case for a general QW. A complete description includes density operator of a QW, which includes both probabilities (diagonal elements) and coherences (offdiagonal elements) in the basis of $\mathpzc{H}_{\,\mathrm{w}}$. To take into account this point, we thus define a more powerful measure of quantumness as the minimum relative entropy of $\varrho_{\textsc{qw}}$ with respect to the set of \textit{classical} states,
\begin{equation}
\label{qu rho}
\mathpzc{Q}(\tau)=\min_{\varrho_{\textsc{rw}}(\tau)} \textit{S}\big(\varrho_{\textsc{qw}}(\tau)\|\varrho_{\textsc{rw}}(\tau)\big),
\end{equation}
where $\textit{S}(\varrho\Vert\sigma)=\mathrm{Tr}[\varrho(\log\varrho-\log\sigma)]$ is the quantum relative entropy \cite{book:Nielsen}. Here the minimum is taken over all density operators $\varrho_{\textsc{rw}}(\tau)$ associated with classical RWs at time $\tau$. These states have a diagonal representation in the position basis with position distribution $\textit{P}_{\textsc{rw}}(\textit{x},\tau)$ as its diagonal elements,
\begin{align}
\label{RhoRWt}
\varrho_{\textsc{rw}}(\tau)=\textstyle{\sum_{\textit{x}=-\tau}^{\tau}} \textit{P}_{\textsc{rw}}(\textit{x},\tau)\project{\textit{x}}.
\end{align}
Such a state can be obtained, e.g., as a result of a QW driven by a noisy quantum coin. In Eq. \eqref{QW and env evo } we choose the contraction noise $\mathpzc{N}(\cdot)=\sum_{i=1}^{4}\textit{K}_{i}\cdot \textit{K}^{\dag}_{i}$ with the Kraus operators $\textit{K}_1=\sqrt{\textit{p}_{+}}\project +$, $\textit{K}_2=\sqrt{\textit{p}_{+}}\ket +\bra -$, $\textit{K}_3=\sqrt{\textit{p}_{-}}\ket -\bra +$, $\textit{K}_4=\sqrt{\textit{p}_{-}}\project -$. This noise channel yields $\mathpzc{N}(\xi)=\textit{p}_{+}\project + + \textit{p}_{-}\project -$ for any input state $\xi$, and when $\mathpzc{N}\otimes \mathpzc{I}$ acts on coin-walker states it removes all correlations and gives a factorized state. Now, with this quantum channel as the evolution $\mathpzc{N}$ and choosing $\textit{U}=\mathbbmss{I}$ or $\textit{U}=\sigma_{\textit{z}}$ (the Pauli matrix $\sigma_{\textit{z}}=|+\rangle\langle +|-|-\rangle\langle -|$) as the quantum coin operator, we find the quantum walker in the classical state given by Eq. \eqref{RhoRWt} \cite{SupplementalMaterial}. We remark that one may devise other methods for assigning quantum states to an RW \cite{szedegy-}. However, our approach is straightforward, and we do not expect that our main message be modified drastically with such methods.
\begin{theorem}
\label{golden}
For a general QW with the position density operator $\varrho_{\textsc{qw}}(\tau)$, we have
\begin{equation}
\label{TotalQ}
\mathpzc{Q}(\tau)=\textit{Q}(\tau)+\textit{C}(\tau),
\end{equation}
where $\textit{C}(\tau)=\textit{H}\big(\textit{P}_{\textsc{qw}}(\tau)\big)-\textit{S}\big(\varrho_{\textsc{qw}}(\tau)\big)$ is the walker's coherence measured by the relative entropy of coherence with respect to the position basis \cite{pleniocoherence2014,q-coherence}. Moreover, the RW which satisfies the minimization in $\mathpzc{Q}$ is the one which minimizes $\textit{Q}$.
\end{theorem}
\begin{figure}[tp]
\includegraphics[scale=0.33]{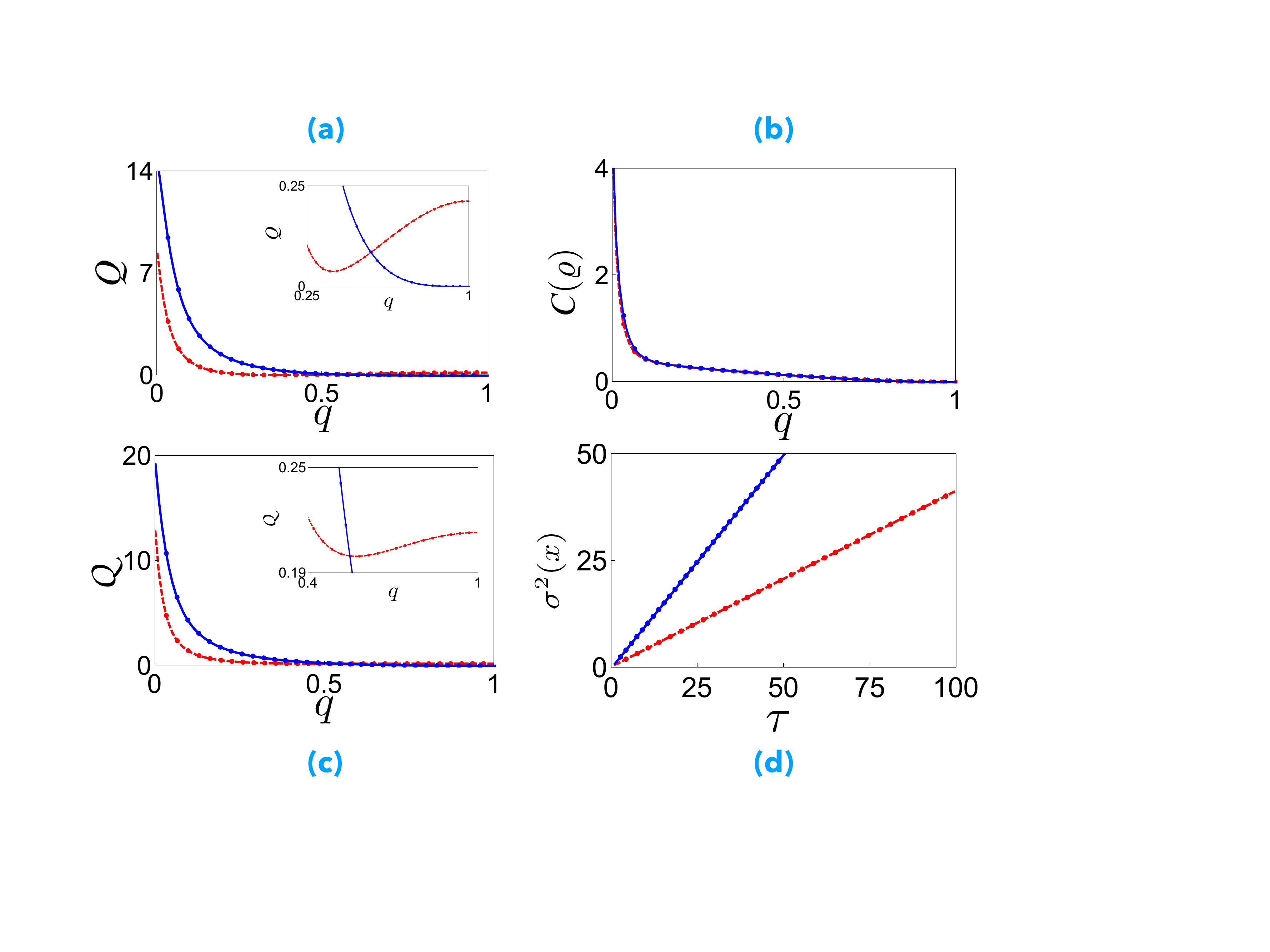}
\caption{(color online). Plots (a), (b), and (c) show $\textit{Q}$, $\textit{C}$, and $\mathpzc{Q}$, respectively, in terms of $q$ at $\tau=100$. Plot (d) shows the variance vs. $\tau$ with $\textit{q}=1$ (note the linear behaviors). The blue curves are for the Hadamard coin and the red dashed curves are for the coin introduced in Eq. \eqref{Uc} with $(\alpha,\beta,\theta)=(0,0,1)$, and here the coin is initially in $\ket +$. The insets show that even for strong decoherence ($\textit{q}\approx 1$) some quantumness remains in the case of the second coin.}
\label{dephasing}
\end{figure}
For a proof see Ref. \cite{ SupplementalMaterial}. Equation \eqref{TotalQ} shows that  $\mathpzc{Q}(\tau)$ captures two kinds of quantumness of a QW; the first quantumness, $\textit{Q}(\tau)$, arises from deviation of the diagonal elements of the walker's density matrix from the probability distribution of classical RWs, and the second quantumness, $\textit{C}(\tau)$, arises from offdiagonal entries of the density matrix of the QW. Hence, this measure of quantumness is in this sense finer than the variance $\sigma^{2}_{\textsc{qw}}$. In fact, there exist QWs that are identified as RW (``classical") by the variance $\sigma^{2}_{\textsc{qw}}$, whereas the quantumness measure $\mathpzc{Q}$ still detects some quantumness existing therein. For example, position variance of all QWs in the presence of unital decaying noise on the coin exhibits $O(\tau)$ asymptotic behavior \cite{brunprl2003}. Consider the noise $\mathpzc{N}$ with $\textit{K}_1=\sqrt{\textit{q}}\project +$, $\textit{K}_2=\sqrt{\textit{q}}\project -$, and $\textit{K}_3=\sqrt{1-\textit{q}}\mathbbmss{I}$. In this case a partial revival occurs for both $\textit{Q}(\tau)$ and $\mathpzc{Q}(\tau)$ after their initial decay even for strong noise values---see Fig. \ref{dephasing}. 

It is in order to make two remarks. (i) Although in this example the final increase in the total quantumness $\mathpzc{Q}$ (inset of plot (c) in Fig. \ref{dephasing}) is because of the increase in the quantumness $\textit{Q}$ (which is based on the probability distributions) while the coherence $\textit{C}$ decays monotonically, in other noise types such as amplitude damping or depolarizing (which is a unital decaying map), quantum coherence $\textit{C}$ also exhibits revival. As demonstrated in Fig. \ref{dephasing}, the variance is linear in both cases for $\textit{q}=1$ at $\tau=100$. Thus, although both walks are classified as classical by their variances $\sigma^{2}$, $\textit{Q}$ indicates only the Hadamard QW as genuinely classical. (ii) The Gaussian reference of a given QW with unital decaying noise can be determined straightforwardly because the first moment of the position $\langle \textit{x}\rangle$ in such cases tends to a constant value \cite{brunprl2003}. For a Hadamard coin with $\textit{q}=1$, this constant vanishes independently of the coin initial state \cite{brunpra2003}, saturating the bound \eqref{upperbound}. 

\begin{figure}[tp]
\includegraphics[scale=0.13]{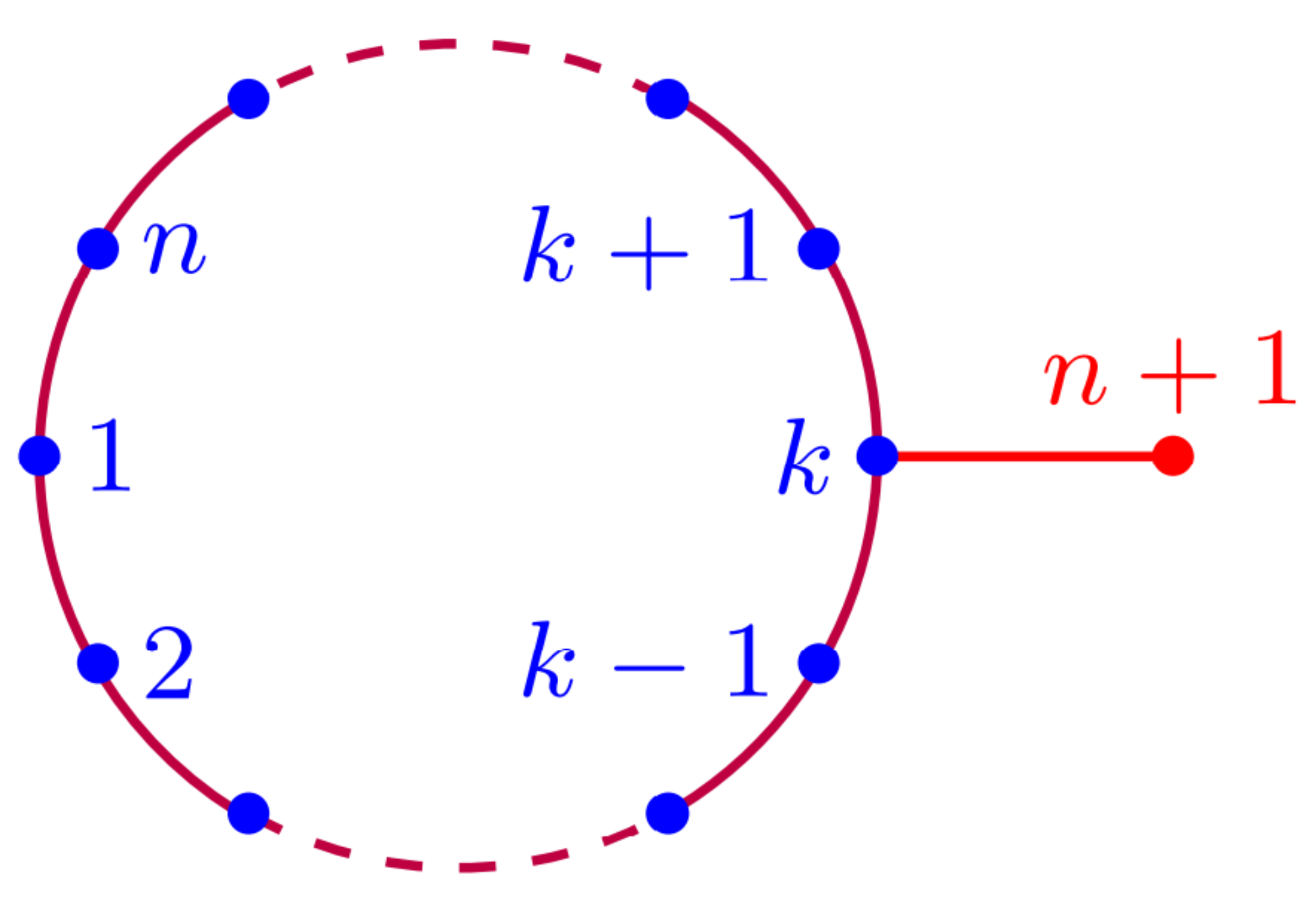}
\caption{A loop with $\textit{n}$ sites and a sink potential acting at site $\textit{k}$, which induces transfer to site $\textit{n}+1$.}
\label{loop}
\end{figure}

\textit{An application.---}Revivals in quantumness of QW may partly explain enhancement of quantum transport in the presence of noise \cite{mohseni,pleniosuasana,chandrashekar2,chin}. Consider, for example, a system consisting of $\textit{n}$ sites in a closed loop and let the system be connected to the environment (site $\textit{n}+1$) via a particular site, say, $\textit{k}$---Fig. \ref{loop}. Discrete QW can be adopted to describe dynamics of the system \cite{chandrashekar2,liu}. The walker either remains in the loop and walks clockwise or counterclockwise (with a probability determined by a quantum coin), or it leaves the loop and never returns. To model this, we extend the system by adding the environment as the extra site $\textit{n}+1$. The walk operator is then the same as Eq. (\ref{qw evo ope}) with the extra properties $\textit{M}_{+} |\textit{n}\rangle=|1\rangle$, $\textit{M}_{+} \ket{\textit{n}+1}=\ket{\textit{n}+1}$, and $\textit{M}_{-}=\textit{M}^{\dag}_{+}$ for a loop with $\textit{n}$ sites. An example of the system-environment interaction is a sink potential assumed at site $\textit{k}$ causing the walker to leave or remains in the loop with the respective probabilities $r$ and $1-r$ \cite{chandrashekar2}. This sink evolution can be simulated by a quantum channel, e.g., $\mathpzc{S}(\cdot)=\sum_{i=1}^{2}\textit{K}_{i}\cdot \textit{K}^{\dag}_{i}$ with $\textit{K}_1=\mathbbmss{I}\otimes\big(\sum_{\textit{x}\neq \textit{k}}|\textit{x}\rangle \langle \textit{x}| +\sqrt{1-\textit{r}}|\textit{k}\rangle \langle \textit{k}|\big)$ and $\textit{K}_2=\mathbbmss{I} \otimes\sqrt{\textit{r}}\ket {\textit{n}+1}\langle \textit{k}|$ acting on the coin-walker state at each step after the action of the unperturbed QW operation $\mathpzc{W}$ (\ref{qw evo ope}). This leads to decay in the probability of finding the walker in the loop. Quantum transport efficiency at time $\tau$ can be quantified by the amount of this decay \cite{chandrashekar2}
\begin{equation}
\eta_{\textsc{qw}}(\tau)=1-\textstyle{\sum_{\textit{x}=1}^{\textit{n}}} \textit{P}_{\textsc{qw}}(\textit{x},\tau)=\textit{P}_{\textsc{qw}}(\textit{n}+1,\tau).
\label{TE}
\end{equation}

The classical version of this transport can be obtained by proper substitution of RW rather than QW. Here the classical walker moves clockwise or counterclockwise, respectively, with probabilities $\textit{p}_{\pm}$ whenever it is in the loop ($\textit{x}\neq \textit{n}+1$), and in case it leaves the loop for the sink site $\textit{n}+1$, it remains there. At each step, the RW is followed by the $(\textit{n}+1)\times (\textit{n}+1)$ classical sink matrix $\textit{S}_{\textit{k}}$, defined as $[\textit{S}_{\textit{k}}]_{\textit{ij}}=\delta_{\textit{ij}}+\textit{r}\delta_{\textit{i}\,\textit{n}+1}\delta_{\textit{jk}}- \textit{r}\delta_{\textit{ik}}\delta_{\textit{jk}}$. RW and sink operators together construct a stochastic transition matrix for the process applied on the $(\textit{n}+1)$-dimensional probability vector for finding the walker at site $1,\ldots, \textit{n}+1$, at each step of the walk. Classical transport efficiency can then be quantified by $\eta_{\textsc{rw}}(\tau)=1-\sum_{\textit{x}=1}^{\textit{n}} \textit{P}_{\textsc{rw}}(\textit{x},\tau)=\textit{P}_{\textsc{rw}}(\textit{n}+1,\tau)$.

\begin{figure}[tp]
\includegraphics[scale=0.325]{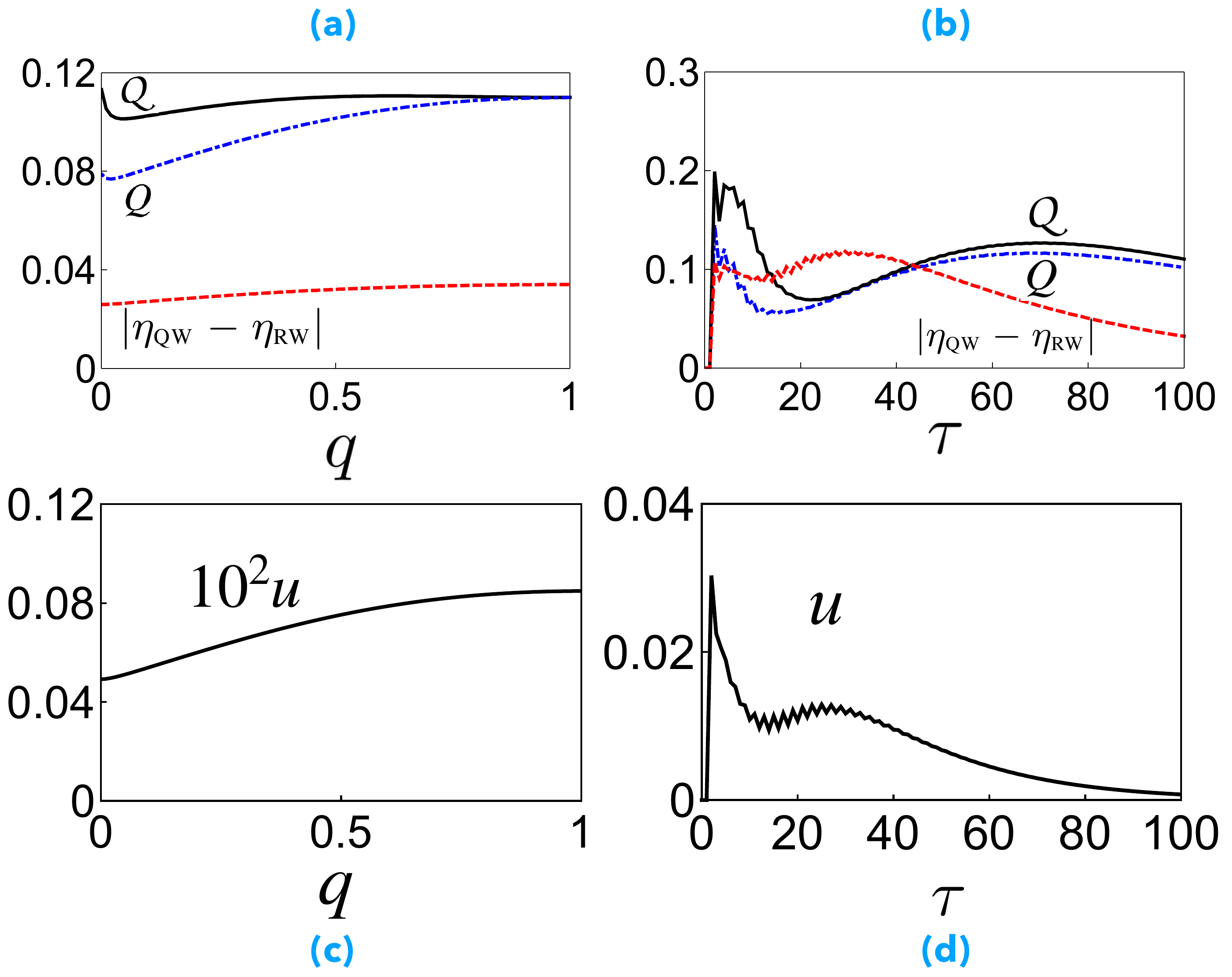}
\caption{For a loop with $\textit{n}=9$, $\textit{k}=3$, and the coin initially in $\ket +$ ($\eta=0$) and characterized by $(\alpha,\beta,\theta)=(0,0,1)$ in Eq. \eqref{Uc}, plot (a) shows $\textit{Q}$, $\mathpzc{Q}$, and $|\eta_{\textsc{qw}}-\eta_{\textsc{rw}}|$ vs. $\textit{q}$ at $\tau=100$. In plot (b) these quantities are depicted vs. $\tau$ at $\textit{q}=1/2$. Plots (c) and (d) represent $\textit{u}$ for the same conditions as plots (a) and (b), respectively, all in $\textit{r}=1$.}
\label{fig:3-all}
\end{figure}

Since quantumness of QW contains all quantum aspects of the walker, especially in the sense of $\mathpzc{Q}$, it should describe the deviation of  $\eta_{\textsc{qw}}$ from the  classical counterpart, i.e., $|\eta_{\textsc{qw}}-\eta_{\textsc{rw}}|$. Figure \ref{fig:3-all} shows the behavior of the quantumness measures $\mathpzc{Q}$, $\textit{Q}$, and $|\eta_{\textsc{qw}}-\eta_{\textsc{rw}}|$ in the presence of the described decoherence noise, from which one can see a relative agreement between these measures. In the trivial case of no sink ($\textit{r}=0$), although the system may have quantum properties, there is no transport in the system (whether classical or quantum) and hence the deviation of the quantum transport from the classical one vanishes too. This is not surprising because quantumness of QW detects total quantum properties of the system whereas quantum transport (in the sense discussed before) is simply a witness for a specific kind of quantumness. In fact, using $\ln \textit{x}\leqslant \textit{x}-1$, one can see 
\begin{equation}
\mathpzc{Q}\geqslant \textit{Q}\geqslant \textit{u}\geqslant 0,
\end{equation}
where $\textit{u}=(\eta_{\textsc{rw}}-\eta_{\textsc{qw}})/\log 2+\eta_{\textsc{qw}}\log (\eta_{\textsc{qw}} / \eta_{\textsc{rw}})$. This lower bound itself may be considered as a measure of distinction between $\eta_{\textsc{qw}}$ and $\eta_{\textsc{rw}}$, because the last inequality is saturated iff $\eta_{\textsc{qw}}=\eta_{\textsc{rw}}$. Figure \ref{fig:3-all} shows that when $\eta_{\textsc{qw}}$ differs from $\eta_{\textsc{rw}}$, the quantumness is nonzero; that is, despite noise induced by the environment, still some quantumness survives. This may suggest that high efficiencies for quantum transport in the presence of noise (e.g., in photosynthetic energy transfer \cite{mohseni}) are related to the existence of quantumness. 

\textit{Summary and discussion.---}We have defined two computable measures of quantumness for quantum walks. Our first measure is based on the difference between the position probability distribution of a quantum walk and that of all (classical) random walks, measured by relative entropy. We have performed the optimization within this definition and showed that the nearest reference random walk to a quantum walk is the one with the same first moment for the position, if the walker was initially localized (extension to other initial conditions is straightforward). In addition, we have introduced a more powerful measure by comparing the density matrix of a quantum walk with that of all classical random walks, characterized by their quantum relative entropy. As an important application of our measures, we have shown that in quantum transport phenomena a relatively high transport efficiency accompanies the existence of some quantumness in the system. Along this line, and in addition to their intrinsic importance, our measures may significantly improve our understanding about whether and how quantumness can yield higher performance of some quantum-walk-based algorithms and tasks. Such understanding and tools can in turn offer novel ways to better harness quantumness as a resource for quantum technologies. 

We end this paper by briefly remarking on possible extensions of our methods. (i) One could have defined quantumness by using other distance measures (although literally relative entropy is not a distance). For example, consider $\textit{Q}=\textit{D}\big(\textit{P}_{\textsc{qw}},\textit{P}_{\textsc{rw}}\big)=\sum_{\textit{x}} \vert \textit{P}_{\textsc{qw}}(\textit{x},\tau)-\textit{P}_{\textsc{rw}}(\textit{x},\tau)\vert$. Such a choice of distance again results in $\mathpzc{Q}(\tau)=\textit{Q}(\tau)+\textit{C}(\tau)$ for the total quantumness, where now $\textit{C}(\tau)$ is the walker's coherence quantified by the $\ell_1$-norm of coherence with respect to the position basis \cite{pleniocoherence2014}. (ii) Note that our results are for the case of initially localized walker. For more general cases another natural protocol is to compare a quantum walk with respect to all random walks in which the walker starts from the positions where the quantum walker has a nonvanishing initial probability (i.e., substituting $\{\textit{P}_{\textsc{rw}}(\textit{x},\tau)\}_{\textit{x}}$ with $\{ \sum_{\textit{x}_{0}}\textit{P}_{\textsc{qw}}(\textit{x}_{0},0)\,\textit{P}_{\textsc{rw}}(\textit{x},\tau; \textit{x}_{0},0)\}_{\textit{x}}$ in Eq. (\ref{p rho})). This option may suit better when we are interested to see how a quantum walk changes if we substitute quantum dynamics with classical dynamics. (iii) One may find a closer classical random walk to a given quantum walk if we relax the condition of equality between the steps of the random walk and the quantum walk and minimize Eq. \eqref{p rho} with respect to classical walker's steps, i.e., $\textit{Q}(\tau)=\min_{\tau'}\min_{\textit{P}_{\textsc{rw}}(\tau')}\textit{D} \big(\textit{P}_{\textsc{qw}}(\tau)\Vert \textit{P}_{\textsc{rw}}(\tau')\big)$. In this case, since at long-time limit the probability distribution $\textit{P}_{\textsc{qw}}$ spreads uniformly in $[-\tau\cos\theta,\tau\cos\theta]$ \cite{Nayak2000,laflamme2008}, the restriction $\tau'\geqslant\tau\cos\theta$ will be necessary to avoid divergence of the relative entropy. Quantifying quantumness in this manner, the usual non-Gaussianity measure is obtained at large-time limit (where the Gaussian reference possesses the same first and second moments with the ones of the distribution of interest).

\textit{Acknowledgments.---}This work was partially supported by Ferdowsi University of Mashhad under Grant No. 3/45469 
and Sharif University of Technology's Office of Vice President for Research.

\newpage
\begin{widetext}
\appendix
\section{Supplemental Materials}
\label{appendix}
In the following we provide proofs for the theorems stated in the main text.

\subsection{I. Analytic expression for the walker's state [Eq. \eqref{ParticleState}]}
\label{appendixa}

 In this section we provide a new method and present a closed relation for the  $\ket{\Psi_{\pm}(\tau)}$. Suppose that the walker starts walking at origin $\ket{0}$ and the initial state of the coin is described by the pure state $\ket{\phi_0}$. We obtain
\begin{align}
\ket{\Psi_{\pm}(\tau)}=& \textstyle{\sum_{\textit{k}_1=0}^1} \cdots\sum_{\textit{k}_{\tau-1}=0}^1 \bra{\pm} \textit{U}  P_{\textit{k}_{\tau-1}}\textit{U} P_{\textit{k}_{\tau-2}} \textit{U}\ldots \textit{U}  P_{\textit{k}_1}\textit{U} \ket{\phi_0}\big|\textstyle{\sum_{i=1}^{\tau-1}} (-1)^{\textit{k}_i} \pm 1\big\rangle, 
\end{align}
where $P_{0}=|+\rangle\langle +|$ and $P_{1}=|-\rangle\langle -|$. These equations can be recast as
\begin{align}
\ket{\Psi_{\pm}(\tau)}=& \textstyle{\sum_{\textit{N}_{+}=0}^{\tau-1}} \bra{\pm}\textit{R}_{\textit{N}_{+}}\ket{\phi_0}\ket{2\textit{N}_{+}-\tau+1\pm1}, 
\end{align}
where $\textit{R}_{\textit{N}_{+}}$s ($\textit{N}_{+}\in\{0,1,\ldots, \tau-1\}$) are $2\times 2$ matrices acting on $\mathpzc{H}_{\,\mathrm{c}}$ and are defined by
\begin{align}
\label{PNR}
\textit{R}_{\textit{N}_{+}} = \textstyle{\sum_{\textit{k}_1,\ldots, \textit{k}_{\tau-1}}^\prime} \textit{U}  P_{\textit{k}_{\tau-1}}\textit{U} P_{\textit{k}_{\tau-2}}\textit{U}\ldots \textit{U}P_{\textit{k}_1}\textit{U}.
\end{align}
Here $\sum^\prime$ is restricted to those values of $\textit{k}_i$ ($i\in\{1,\ldots,\tau-1\}$) which satisfy the condition $\sum_{i=1}^{\tau-1} \textit{k}_i=\textit{N}_{-}=\tau-1-\textit{N}_{+}$. In the basis $\{\ket \pm\}$, $\textit{R}_{\textit{N}_{+}}$ has the following matrix elements:
\begin{equation}\label{pi n r}
[\textit{R}_{\textit{N}_{+}}]_{\textit{k}_{\tau} \textit{k}_0} = \textstyle{\sum_{\textit{k}_1,\ldots, \textit{k}_{\tau-1}}^\prime} \textit{U}_{\textit{k}_{\tau} \textit{k}_{\tau-1}}\textit{U}_{\textit{k}_{\tau-1}\textit{k}_{\tau-2}}\ldots \textit{U}_{\textit{k}_1\textit{k}_0},
\end{equation}
with $\textit{U}_{ij}=[\textit{U}]_{ij}$. For the extreme cases $\textit{N}_{+}=0$ or $\textit{N}_{+}=\tau-1$, which indicate that the walker moves only to the left or to the right, respectively, Eq. \eqref{pi n r} yields
\begin{align}
[\textit{R}_{0}]_{\textit{k}_{\tau} \textit{k}_0}=&\textit{U}_{\textit{k}_{\tau} 1}\textit{U}_{11}^{\tau-2}\textit{U}_{1\textit{k}_0},\\
[\textit{R}_{\tau-1}]_{\textit{k}_{\tau} \textit{k}_0}=&\textit{U}_{\textit{k}_{\tau} 0} \textit{U}_{00}^{\tau-2}\textit{U}_{0\textit{k}_0}.
\end{align}
For the other values $\textit{N}_{+}\in\{1,2,\ldots,\tau-2\}$ we obtain
\begin{align}
[\textit{R}_{\textit{N}_{+}}]_{\textit{k}_{\tau} \textit{k}_0}=& \textit{U}_{00}^{\textit{N}_{+}-2} \textit{U}_{11}^{\textit{N}_{-}-2} \Big[ {}_{2}\textit{F}_{1}(1-\textit{N}_{+},1-\textit{N}_{-};1;\textit{z})\,\textit{U}_{\textit{k}_{\tau}0}\textit{U}_{00} \textit{U}_{11} \textit{U}_{01}\textit{U}_{1\textit{k}_0}\nonumber\\
&+ {}_{2}\textit{F}_{1}(1-\textit{N}_{+},1-\textit{N}_{-};1;\textit{z})\,\textit{U}_{\textit{k}_{\tau}1}\textit{U}_{00}\textit{U}_{11}\textit{U}_{10}\textit{U}_{0\textit{k}_0} \nonumber\\ 
&+ {}_{2}\textit{F}_{1}(2-\textit{N}_{+},1-\textit{N}_{-};2;\textit{z})\,\textit{U}_{\textit{k}_{\tau}0} \textit{U}_{11}\textit{U}_{01}\textit{U}_{10}\textit{U}_{0\textit{k}_0}(\textit{N}_{+}-1)\nonumber\\ 
&+ {}_{2}\textit{F}_{1}(1-\textit{N}_{+},2-\textit{N}_{-};2;\textit{z})\,\textit{U}_{\textit{k}_{\tau}1}\textit{U}_{00}\textit{U}_{01}\textit{U}_{10}\textit{U}_{1\textit{k}_0}(\textit{N}_{-}-1)\Big], \label{pi n r2}
\end{align}
where $\textit{z}=(\textit{U}_{01}\textit{U}_{10})/(\textit{U}_{00}\textit{U}_{11})$ and ${}_{2}F_{1}(\textit{a},\textit{b};\textit{c};\textit{z})$ is the hypergeometric function defined as \cite{book:Arfken}
\begin{align}
{}_{2}\textit{F}_{1}(\textit{a},\textit{b};\textit{c};\textit{z})=1+\frac{\textit{ab}}{\textit{c}}\frac{\textit{z}}{1!}+\frac{\textit{a}(\textit{a}+1)\textit{b}(\textit{b}+1)}{\textit{c}(\textit{c}+1)}\frac{\textit{z}^2}{2!}+\ldots,\quad \textit{c}\notin\{0,-1,-2,\ldots\}.
\end{align} 

A generalization of the above discussion to the case where the walk is driven by a coin initially in a mixed state is straightforward. To prove Eq. \eqref{pi n r2} we write matrix elements of $\textit{R}_{\textit{N}_{+}}$ given by Eq. \eqref{PNR} as ($\textit{N}_{+}\in\{0,1,\ldots,\textit{k}_{\tau-1}\}$)
\begin{align}
\label{proof of pi n r}
[\textit{R}_{\textit{N}_{+}}]_{\textit{k}_{\tau}\textit{k}_0} = \textstyle{\sum_{\textit{k}_1,\ldots, \textit{k}_{\tau-1}}^\prime} \bra{\textit{k}_{\tau}}\textit{U} P_{\textit{k}_{\tau-1}}\textit{U} P_{\textit{k}_{\tau-2}}\textit{U}\ldots UP_{\textit{k}_1}\textit{U}\ket{\textit{k}_0},
\end{align}
recalling that $\sum^\prime$ is restricted to those values of $\textit{k}_i$ ($i\in\{1,\ldots,\tau-1\})$ which satisfy the condition $\sum_{i=1}^{\tau-1} \textit{k}_i=\textit{N}_{-}=\tau-1-\textit{N}_{+}$.
The above summation can be divided, in general, into four classes just by considering all possible choices for the first and last projectors, i.e., the pair $\{P_{\textit{k}_{\tau-1}}, P_{\textit{k}_1}\}$. Indeed, the pair takes any value of $\{P_{0}, P_{1}\}$, $\{P_{1}, P_{0}\}$, $\{P_{0}, P_{0}\}$, and $\{P_{1}, P_{1}\}$. Depending on to what class the terms of the summation belongs, one can rewrite the above equation as
\begin{align}
\nonumber
[\textit{R}_{\textit{N}_{+}}]_{\textit{k}_{\tau}\textit{k}_0}=& \textstyle{\sum_{\textit{k}_2,\ldots, \textit{k}_{\tau-2}}^\prime} \bra{\textit{k}_{\tau}}\textit{U} P_{0}\textit{U} P_{\textit{k}_{\tau-2}}U\ldots \textit{U}P_{1} \textit{U}\ket{\textit{k}_0}\\ 
\nonumber & + \textstyle{\sum_{\textit{k}_2,\ldots, \textit{k}_{\tau-2}}^\prime} \bra{\textit{k}_{\tau}}\textit{U} P_{1}\textit{U} P_{\textit{k}_{\tau-2}}\textit{U}\ldots \textit{U}P_{0}\textit{U} \ket{\textit{k}_0}\\ 
\nonumber &+ \textstyle{\sum_{\textit{k}_2,\ldots, \textit{k}_{\tau-2}}^\prime}\bra{\textit{k}_{\tau}}\textit{U} P_{0}\textit{U} P_{\textit{k}_{\tau-2}}\textit{U}\ldots \textit{U}P_{0} \textit{U}\ket{\textit{k}_0}\\ 
\nonumber &+ \textstyle{\sum_{\textit{k}_2,\ldots, \textit{k}_{\tau-2}}^\prime} \bra{\textit{k}_{\tau}}\textit{U} P_{1}\textit{U} P_{\textit{k}_{\tau-2}}\textit{U}\ldots \textit{U}P_{1}\textit{U}\ket{\textit{k}_0}.
\end{align}
The restriction on the summations now changes for various summations. For example, for the first one $\sum^\prime$ is restricted to those values $0,1$ of $\textit{k}_i$ ($i\in\{2,\ldots,\tau-2\})$ which satisfy the condition $\sum_{i=2}^{\tau-2} \textit{k}_i=\textit{N}_{-}-1=\tau-2-\textit{N}_{+}$. In what follows we focus only on the first summation---the proof of the others are straightforward. In this case we should sum all allowed compositions which start with $(\textit{U} P_0)$ and end in $(\textit{U} P_1)\textit{U}$. Such compositions are of the form
\begin{align}
\underbrace{(\textit{U}P_0)\ldots(\textit{U}P_0)}_{a_1}\underbrace{(\textit{U}P_1)\ldots(\textit{U}P_1)}_{b_1} \ldots \underbrace{(\textit{U}P_0)\ldots(\textit{U}P_0)}_{a_{\textit{N}}}\underbrace{(\textit{U}P_1)\ldots(\textit{U}P_1)}_{b_{\textit{N}}}\textit{U},
\end{align}
where $a_i$ and $b_i$ represent the number of $(\textit{U}P_0)$s and $(\textit{U}P_1)$s in block $i$, respectively, so the restrictions $\sum_{i=1}^{\textit{N}}a_i=\textit{N}_{+}$ and $\sum_{i=1}^{\textit{N}} b_i=\textit{N}_{-}$ exist and obviously $\textit{N}\in\big\{1, 2, \ldots, \min\{\textit{N}_{+},\textit{N}_{-}\}\big\}$. Thus for the first summation we have
\begin{align} 
& \textstyle{\sum_{\textit{k}_2,\ldots, \textit{k}_{\tau-2}}^\prime} \bra{\textit{k}_{\tau}}\textit{U} P_{0}\textit{U} P_{\textit{k}_{\tau-2}}U\ldots \textit{U}P_{1}{U}\ket{\textit{k}_0} \qquad\qquad \qquad \qquad\nonumber\\
&= \textstyle{\sum_{\textit{N}}} \textstyle{\sum_{a_1,\ldots, a_{\textit{N}}\atop {b_1,\ldots, b_{\textit{N}}}}^\prime} \bra{\textit{k}_{\tau}}(\textit{U}P_0)^{a_1}(\textit{U}P_1)^{b_1}\ldots(\textit{U}P_0)^{a_{\textit{N}}}(\textit{U}P_1)^{b_{\textit{N}}}\textit{U}\ket{\textit{k}_0} \nonumber\\
&= \textstyle{\sum_{\textit{N}}} \textstyle{\sum_{a_1,\ldots, a_{\textit{N}}\atop {b_1,\ldots, b_{\textit{N}}}}^\prime} \textit{U}_{\textit{k}_{\tau}0} \textit{U}_{00}^{a_1-1} \textit{U}_{01} \textit{U}_{11}^{b_1-1} \textit{U}_{10}\ldots \textit{U}_{00}^{a_{\textit{N}}-1} \textit{U}_{01}\textit{U}_{11}^{b_{\textit{N}}-1} \textit{U}_{1\textit{k}_0} \nonumber\\
&= \textstyle{\sum_{\textit{N}}} C_{\textit{N}}^{\textit{N}_{+}}C_{\textit{N}}^{\textit{N}_{-}} \textit{U}_{\textit{k}_{\tau}0} \textit{U}_{00}^{\textit{N}_{+}-N} \textit{U}_{11}^{\textit{N}_{-}-\textit{N}} \textit{U}_{01}^{\textit{N}} \textit{U}_{10}^{N-1} \textit{U}_{1\textit{k}_0}. \qquad \qquad
\end{align}
In writing the last equation we have incorporated the restrictions on $a_i$ and $b_i$, and also $C_{\textit{N}}^{\textit{N}_{+}}$ and $C_{\textit{N}}^{\textit{N}_{-}}$ count all possible combinations satisfying these restrictions, respectively, for a fixed $\textit{N}$. Thus our problem now reduces to enumerating possible ways of distributing $\textit{N}_{+}$ matrices $(\textit{U}P_0)$ in $\textit{N}$ blocks (and the same for $\textit{N}_{-}$ matrices $(\textit{U}P_1)$), in such a way that each of these blocks contains at least $1$ matrix as $\textit{N}$ is fixed. This is equivalent to the problem of distributing $\textit{N}_{+}$ indistinguishable balls in $\textit{N}$ boxes in a way that there remains no empty box. The number of all possible combinations for such a problem is $\frac{(\textit{N}_{+}-1)!}{(\textit{N}-1)!(\textit{N}_{+}-\textit{N})!}$, so $C_{\textit{N}}^{\textit{N}_{+}}=\frac{(\textit{N}_{+}-1)!}{(N-1)!(\textit{N}_{+}-\textit{N})!}$ and $C_{\textit{N}}^{\textit{N}_{-}}=\frac{(\textit{N}_{-}-1)!}{(\textit{N}-1)!(\textit{N}_{-}-\textit{N})!}$. Using this, the first summation reduces to
\begin{align}
\textstyle{\sum_{\textit{k}_2,\ldots, \textit{k}_{\tau-2}}^\prime} \bra{\textit{k}_{\tau}}\textit{U} P_{0}\textit{U} P_{\textit{k}_{\tau-2}}U\ldots UP_{1}{U}\ket{\textit{k}_0} & = \textstyle{\sum_{\textit{N}}} C_{\textit{N}}^{\textit{N}_{+}}C_{\textit{N}}^{\textit{N}_{-}} \textit{U}_{\textit{k}_{\tau}0}\textit{U}_{00}^{\textit{N}_{+}-\textit{N}} \textit{U}_{11}^{\textit{N}_{-}-\textit{N}} \textit{U}_{01}^{\textit{N}} \textit{U}_{10}^{\textit{N}-1} \textit{U}_{1\textit{k}_0} \nonumber\\
&= {}_{2}\textit{F}_{1}(1-\textit{N}_{+},1-\textit{N}_{-};1;\textit{z}) \textit{U}_{\textit{k}_{\tau}0} \textit{U}_{00}^{\textit{N}_{+}-1} \textit{U}_{11}^{\textit{N}_{-}-1} \textit{U}_{01} \textit{U}_{1\textit{k}_0},\nonumber
\end{align}
where in the last line we have used the definition of the hypergeometric function \cite{book:Arfken}.

\subsection{II. Optimal point of Eq. \eqref{p rho} }

Using the binomial distribution $\textit{P}_{\textsc{rw}}(\textit{x},\tau)=\binom{\tau}{(\textit{x}+\tau)/2} \textit{p}_{+}^{(\textit{x} +\tau)/2} \textit{p}_{-}^{(\tau-\textit{x})/2}$ in Eq. \eqref{p rho} we obtain
\begin{align}
\label{ProofL1}
\textit{Q}(\tau)=&-\textit{H}\big(\textit{P}_{\textsc{qw}}(\tau)\big) -\max_{\textit{p}_{+}} \textstyle{\sum_{\textit{x}=-\tau}^{\tau}} \textit{P}_{\textsc{qw}}(\textit{x},\tau)\log \textit{P}_{\textsc{rw}}(\textit{x},\tau) \\ \nonumber
=& -\textit{H}\big(\textit{P}_{\textsc{qw}}(\tau)\big)- \textstyle{\sum_{\textit{x}=\tau}^{\tau}} \textit{P}_{\textsc{qw}}(\textit{x},\tau)\log\frac{\tau!}{(\frac{\textit{x} +\tau}{2})! (\frac{\tau-\textit{x}}{2})!} -\max_{\textit{p}_{+}}\Big\{\log \textit{p}_{+} \textstyle{\sum_{\textit{x}=-\tau}^{\tau}} (1/2)(\textit{x} + \tau)\,\textit{P}_{\textsc{qw}}(\textit{x},\tau) \\ \nonumber
&+\log (1-\textit{p}_{+}) \textstyle{\sum_{\textit{x}=-\tau}^{\tau}} (1/2)(\tau-\textit{x})\textit{P}_{\textsc{qw}}(\textit{x},\tau)\Big\}.
\end{align}
Straightforward calculation shows that the maximum of the last term is attained if the RW is driven by a coin with associated probability $\textit{p}_{+}^{*} = (1/2\tau)\sum_{\textit{x}=-\tau}^{\tau}(\textit{x}+\tau)\,\textit{P}_{\textsc{qw}}(\textit{x},\tau)$. 

\subsection{III. Proof of Theorem \ref{LemmaRef} }

Adopting $\textit{p}_{+}^{*}$ as the optimum value in the second equality of Eq. \eqref{ProofL1}, we obtain
\begin{align*}
\textit{Q}(\tau)=&-\textit{H}\big(\textit{P}_{\textsc{qw}}(\tau)\big)+\tau \textit{H}\big(\textit{p}_{+}^{*}(\tau)\big) - \textstyle{\sum_{\textit{x}=-\tau}^{\tau}} \textit{P}_{\textsc{qw}}(\textit{x},\tau)\log\frac{\tau!\,(1/2)^{\tau} (1/2)^{-\tau}}{[(\tau+\textit{x})/2]!\, [(\tau-\textit{x})/2]!} \\
=&\textit{D}(\textit{P}_{\textsc{qw}}(\tau)\|\textit{P}_{\textsc{rw}}^{\textsc{1/2}}(\tau))-\tau\big(1-\textit{H}(p_{+}^{*}(\tau))\big).
\end{align*}

\subsection{IV. Proof of Theorem \ref{golden} }

We recall that the relative entropy of coherence has the closed form $\textit{C}(\varrho)=\textit{S}(\varrho^{\diag})-\textit{S}(\varrho)$, where $\varrho^{\diag}$ denotes the diagonal state obtained from diagonal elements of $\varrho$ in the fixed basis in which the coherence is defined. In our QW model, $\varrho^{\diag}_{\textsc{qw}}(\tau)$ is given by $\textit{P}_{\textsc{qw}}(\tau)$, thus $\textit{S}(\varrho^{\diag}_{\textsc{qw}}(\tau))=\textit{H}(\textit{P}_{\textsc{qw}}(\tau))$, and we obtain
\begin{align*}
\mathpzc{Q}(\tau)=&\min_{\varrho_{\textsc{rw}}(\tau)}\Big(\Tr[\varrho_{\textsc{qw}}(\tau)\log\varrho_{\textsc{qw}}(\tau)]-\Tr[\varrho_{\textsc{qw}}(\tau)\log\varrho_{\textsc{rw}}(\tau)]\Big)\cr
=&-\textit{S}(\varrho_{\textsc{qw}}(\tau))- \max_{\varrho_{\textsc{rw}}(\tau)}\Tr[\varrho_{\textsc{qw}}(\tau)\log{\varrho_{\textsc{rw}}}(\tau)]\cr
=&\textit{C}(\varrho_{\textsc{qw}}(\tau))- \textit{S}(\varrho_{\textsc{qw}}^{\diag}(\tau)) -\max_p\sum_{\textit{x}=-\tau}^{\tau}\textit{P}_{\textsc{qw}}(\textit{x},\tau)\log{\textit{P}_{\textsc{rw}}(\textit{x},\tau)} \\
=&\textit{C}(\varrho_{\textsc{qw}}(\tau))+\min_{\textit{p}_{+}} \textit{H}\big(\textit{P}_{\textsc{qw}}(\tau)\|\textit{P}_{\textsc{rw}}(\tau)\big) \\
=& \textit{C}(\varrho_{\textsc{qw}}(\tau))+ \textit{Q}(\varrho_{\textsc{qw}}(\tau)).
\end{align*}

 Note that the last line of the proof shows that for both $\textit{Q}(\tau)$ and $\mathpzc{Q}(\tau)$ the optimal $\textit{p}_{+}$ is the same.

\subsection{V. Proof of Eq. (\ref{RhoRWt})}

With this quantum channel as the nonunitary evolution $\mathpzc{N}$ and choosing the identity $\mathbbmss{I}$ or the Pauli matrix $\sigma_{\textit{z}}$ as the unitary evolution $\textit{U}$ of the quantum coin, we find the quantum walker in the classical state given by Eq. \eqref{RhoRWt}. Specifically, after some algebra we obtain the coin-walker state as
\begin{align}
\varrho_{\textsc{qw}}(\tau)=\textstyle{\sum_{\textit{m}=0}^{\tau-1}} \frac{(\tau-1)!\,\textit{p}_{+}^{\textit{m}} \textit{p}_{-}^{\tau-\textit{m}-1}}{\textit{m}!(\tau-\textit{m} -1)!}\big(\textit{p}_{+}\project +\otimes\project{2\textit{m}-\tau+2} +\textit{p}_{-}\project -\otimes\project{2\textit{m}-\tau}\big),
\end{align}
where after tracing out over the coin, we find the quantum walker in the classical state given by Eq. \eqref{RhoRWt}.

\twocolumngrid
\end{widetext}

\begin{thebibliography}{99}

\bibitem{rw-book} J. Rudnick and G. Gaspari, \emph{Elements of the Random Walk: An introduction for Advanced Students and Researchers} (Cambridge University Press, Cambridge, 2004).


\bibitem{book:reichl} L. E. Reichl, \emph{A Modern Course in Statistical Physics} (John Wiley \& Sons, New York, 1998).

\bibitem {wright}S. Wright, \href{http://www.genetics.org/content/16/2/97}{Genetics \textbf{16}, 97 (1931)}.

\bibitem {chandrasekhar} S. Chandrasekhar, \href{https://journals.aps.org/rmp/abstract/10.1103/RevModPhys.15.1}{Rev. Mod. Phys. \textbf{15}, 1 (1943)}.

\bibitem{chem} G. H. Weiss and R. J. Rubin, in \href{https://doi.org/10.1002/9780470142769.ch5}{\textit{Advances in Chemical Physics, Vol. 52}}, edited by I. Prigogine and S. A. Rice (New York, John Wiley \& Sons, 2007).

\bibitem{chem-2} T. Aquino and M. Dentz, \href{https://doi.org/10.1103/PhysRevLett.119.230601}{Phys. Rev. Lett. \textbf{119}, 230601 (2018)}. 

\bibitem{rw-biology:book} H. C. Berg, \emph{Random Walks in Biology} (Princeton University Press, Princeton, 1993); E. A. Codling, M. J. Plank, and S. Benhamou, \href{http://rsif.royalsocietypublishing.org/content/5/25/813.long}{J. Roy. Soc. Interface \textbf{5}, 813 (2008).}


\bibitem {nosofsky} R. M. Nosofsky and T. J. Palmeri,  \href{http://psycnet.apa.org/record/1997-03612-003}{Psychol. Rev. \textbf{104}, 266 (1997)}.

\bibitem{rw-economics} E. Scalas, \href{https://doi.org/10.1016/j.physa.2005.11.024/}{Physica A \textbf{362}, 225 (2006).}

\bibitem{rw-fractal} A. Blumen, J. Klafter, B. S. White, and G. Zumofen, \href{https://doi.org/10.1103/PhysRevLett.53.1301}{Phys. Rev. Lett. \textbf{53}, 1301 (1984).}

\bibitem{rw-fb} D. Cassai and S. Regina, \href{https://doi.org/10.1103/PhysRevLett.76.2914}{Phys. Rev. Lett. \textbf{76}, 2914 (1996).}

\bibitem{rw-complex} J. D. Noh and H. Rigger, \href{https://doi.org/10.1103/PhysRevLett.92.118701}{Phys. Rev. Lett. \textbf{92}, 118701 (2004)}.

\bibitem {AharonovPRA1993} Y. Aharonov, L. Davidovich, and N. Zagury, \href{https://journals.aps.org/pra/abstract/10.1103/PhysRevA.48.1687}{Phys. Rev. A \textbf{48}, 1687 (1993)}.

\bibitem{farhi contin} E. Farhi and S. Gutmann, \href{https://journals.aps.org/pra/abstract/10.1103/PhysRevA.58.915}{Phys. Rev. A \textbf{58}, 915 (1998)}.

\bibitem{d-aharonov} D. Aharonov, A. Ambainis, J. Kempe, and U. Vazirani, in \href{https://dl.acm.org/citation.cfm?id=380758}{Proceedings of the 33rd ACM Symposium on Theory of Computing (2001), 50.}

\bibitem{ambainis-1d} A. Ambainis, E. Bach, A. Nayak, A. Vishwanath, and J. Watrous, in \href{https://dl.acm.org/citation.cfm?id=380757}{Proceedings of the 33rd ACM Symposium on Theory of Computing (2001), 37.}


\bibitem{qw-book} R. Portugal, \emph{Quantum Walks and Search Algorithms} (Springer, New York, 2013).

\bibitem {aaronson} S. Aaronson and A. Ambainis, \href{http://www.theoryofcomputing.org/articles/v001a004/}{Theory of Computing \textbf{1}, 47 (2005)}.


\bibitem {childs pra2004 1} A. M. Childs and J. Goldstone, \href{https://journals.aps.org/pra/abstract/10.1103/PhysRevA.70.022314}{Phys. Rev. A \textbf{70}, 022314 (2004)}.



\bibitem {shenvi pra2003} N. Shenvi, J. Kempe, and K. B. Whaley, \href{https://journals.aps.org/pra/abstract/10.1103/PhysRevA.67.052307}{Phys. Rev. A \textbf{67}, 052307 (2003)}.


\bibitem {farhi thcom2008} E. Farhi, J. Goldstone, and S. Gutmann, \href{https://theoryofcomputing.org/articles/v004a008/}{Theory of Computing \textbf{4}, 169 (2008)}.



\bibitem{ambainis2003} A. Ambainis, \href{http://www.worldscientific.com/doi/abs/10.1142/S0219749903000383}{Int. J. Quantum Info. \textbf{1}, 507 (2003)}.

\bibitem{childs-qip-qalgorithm} A. M. Childs, E. Farhi, and S. Gutmann, \href{https://link.springer.com/article/10.1023/A:1019609420309}{Quantum Inf. Process. \textbf{1}, 35 (2002).}


\bibitem{childs2009prl} A. M. Childs, \href{https://journals.aps.org/prl/abstract/10.1103/PhysRevLett.102.180501}{Phys. Rev. Lett. \textbf{102}, 180501 (2009)}.

\bibitem{childs-science}A. M. Childs, D. Gosset, and Z. Webb, \href{http://science.sciencemag.org/content/339/6121/791}{Science \textbf{339}, 791 (2013).} 

\bibitem{omar-1} S. Chakraborty, L. Novo, A. Ambainis, and Y. Omar, \href{https://doi.org/10.1103/PhysRevLett.116.100501}{Phys. Rev. Lett. \textbf{116}, 100501 (2016)}.

\bibitem{omar-2} S. Chakraborty, L. Novo, S. Di Giorgio, and Y. Omar, \href{https://doi.org/10.1103/PhysRevLett.119.220503}{Phys. Rev. Lett. \textbf{119}, 220503 (2017)}.

\bibitem{logic-walk} Y. Lahini, G. R. Steinbrecher, A. D. Bookatz, and D. Englund, \href{http://www.nature.com/articles/s41534-017-0050-2/}{njp Quantum Info. \textbf{4}, 2 (2008)}.

\bibitem{science-strong} P. M. Preiss, R. Ma, M. E. Tai, A. Lukin, M. Rispoli, P. Zupancic, Y. Lahini, R. Islam, and M. Greiner, \href{http://science.sciencemag.org/content/347/6227/1229}{Science \textbf{347}, 1229 (2015)}.

\bibitem{sciarrino} T. Giordani, E. Polino, S. Emiliani, A. Suprano, L. Innocenti, H. Majury, L. Marrucci, M. Paternostro, A. Ferraro, N. Spagnolo, and F. Sciarrino, \href{https://arxiv.org/abs/1808.08875}{arXiv:1808.08875 (2018)}.

\bibitem{walk-bec} S. Dadras, A. Gresch, C. Groiseau, S. Wimberger, and G. S. Summy, \href{https://doi.org/10.1103/PhysRevLett.121.070402}{Phys. Rev. Lett. \textbf{121}, 070402 (2018)}.

\bibitem{mulken} O. M\"{u}lken and  A. Blumen, \href{https://doi.org/10.1016/j.physrep.2011.01.002}{Phys. Rep. \textbf{502}, 37 (2011).}

\bibitem {dorner} R. Dorner, J. Goold, and V. Vedral, \href{http://rsfs.royalsocietypublishing.org/content/2/4/522}{Interface Focus  \textbf{2}, 522  (2012)}.

\bibitem {chandrashekar2} C. M. Chandrashekar and T. Busch, \href{https://link.springer.com/article/10.1007/s11128-014-0730-1}{Quantum Inf. Process. \textbf{13}, 1313 (2014)}.

\bibitem{mohseni} M. Mohseni, P. Rebentrost, S. Lloyd, and A. Aspuru-Guzik, \href{http://aip.scitation.org/doi/abs/10.1063/1.3002335}{J. Chem. Phys. \textbf{129}, 174106 (2008)}.

\bibitem {pleniosuasana} M. B. Plenio and S. F. Huelga, \href{http://iopscience.iop.org/article/10.1088/1367-2630/10/11/113019/meta}{New J. Phys.  \textbf{10}, 113019  (2008)}.

\bibitem{chin} A. W. Chin, A. Datta, F. Caruso, S. F. Huelga, and M. B. Plenio, \href{http://iopscience.iop.org/article/10.1088/1367-2630/12/6/065002/meta}{New J. Phys. \textbf{12},  065002 (2010)}.

\bibitem{Panitchayangkoona} G. Panitchayangkoon, D. V. Voronine, D. Abramavicius, J. R. Caram, N. H. C. Lewis, S. Mukamel, and G. S. Engel, \href{http://www.pnas.org/content/108/52/20908.abstract}{Proc. Natl. Acad. Sci. USA \textbf{108}, 20908 (2011)}.

\bibitem{book:Nielsen} M. A. Nielsen and I. L. Chuang, \emph{Quantum Computation and Quantum Information} (Cambridge University Press, Cambridge, 2000).

\bibitem{strength} C. H. Bennett, E. Bernstein, G. Brassard, and U. Vazirani, \href{https://doi.org/10.1137/S0097539796300933/}{SIAM J. Comput. \textbf{26}, 1510 (2006)}.

\bibitem{preskill} J. Preskill, \href{https://arxiv.org/pdf/1203.5813.pdf}{arXiv:1203.5813 (2012)}.


\bibitem{meyer} D. A. Meyer, \href{https://link.springer.com/article/10.1007\%2FBF02199356}{J. Stat. Phys. \textbf{85}, 551 (1996)}.

\bibitem{brunprl2003} T. A. Brun, H. A. Carteret, and A. Ambainis, \href{https://journals.aps.org/prl/abstract/10.1103/PhysRevLett.91.130602}{ Phys. Rev. Lett. \textbf{91}, 130602 (2003)}.

\bibitem{romanelli 2} A. Romanelli, R. Siri, G. Abal, A. Auyuanet, and R. Donangelo, \href{http://www.sciencedirect.com/science/article/pii/S0378437104011422?via\%3Dihub}{ Physica A \textbf{347}, 137 (2005)}.


\bibitem{shapira} D. Shapira, O. Biham, A. J. Bracken, and M. Hackett, \href{https://journals.aps.org/pra/abstract/10.1103/PhysRevA.68.062315}{ Phys. Rev. A \textbf{68}, 062315 (2003)}.



\bibitem{rezakhani} J. B. Stang, A. T. Rezakhani, and B. C. Sanders, \href{http://iopscience.iop.org/article/10.1088/1751-8113/42/17/175304/meta}{J. Phys. A: Math. Theor. \textbf{42}, 175304 (2009)}.

\bibitem{Amin} M. H. S. Amin, P. J. Love, and C. J. S. Truncik, \href{https://journals.aps.org/prl/abstract/10.1103/PhysRevLett.100.060503}{Phys. Rev. Lett. \textbf{100}, 060503 (2008)}.

\bibitem{Kasal} I. Kassal, J. Yuen-Zhou, and S. Rahimi-Keshari, \href{https://pubs.acs.org/doi/10.1021/jz301872b}{J. Phys. Chem. Lett. \textbf{4}, 362 (2013)}.

\bibitem{ekert-papers} I. Sinayskiy, A. Marais, F. Petruccione, and A. Ekert, \href{https://doi.org/10.1103/PhysRevLett.108.020602}{Phys. Rev. Lett. \textbf{108}, 020602 (2012)}.

\bibitem{childs-cmp} A. M. Childs, \href{https://link.springer.com/article/10.1007%2Fs00220-009-0930-1}{Commun. Math. Phys. \textbf{294}, 581 (2010)}.


\bibitem{montero pra} M. Montero, \href{https://journals.aps.org/pra/abstract/10.1103/PhysRevA.95.062326}{Phys. Rev. A \textbf{95}, 839 (2017)}.

\bibitem {Nayak2000} A. Nayak and A. Vishwanath, \href{https://arxiv.org/abs/quant-ph/0010117}{arXiv:quant-ph/0010117 (2000)}.

\bibitem{SupplementalMaterial} See the Supplemental Materials for the proofs.



\bibitem{lopez} C. C. L\'{o}pez and J. P. Paz, \href{https://journals.aps.org/pra/abstract/10.1103/PhysRevA.68.052305}{ Phys. Rev. A \textbf{68}, 052305 (2003)}.


\bibitem{annabestani} M. Annabestani, S. J. Akhtarshenas, and M. R. Abolhassani, \href{https://journals.aps.org/pra/abstract/10.1103/PhysRevA.81.032321}{ Phys. Rev. A \textbf{81}, 032321 (2010)}.

\bibitem{vedral97} V. Vedral, M. B. Plenio, M. A. Rippin, and P. L. Knight, \href{https://journals.aps.org/prl/abstract/10.1103/PhysRevLett.78.2275}{ Phys. Rev. Lett. \textbf{78}, 2275 (1997)}.


\bibitem{modi} K. Modi, T. Paterek, W. Son, V. Vedral, and M. Williamson, \href{https://journals.aps.org/prl/abstract/10.1103/PhysRevLett.104.080501}{Phys. Rev. Lett. \textbf{104}, 080501 (2010)}.

\bibitem{pleniocoherence2014} T. Baumgratz, M. Cramer, and M. B. Plenio, \href{https://journals.aps.org/prl/abstract/10.1103/PhysRevLett.113.140401}{ Phys. Rev. Lett. \textbf{113}, 140401 (2014)}.

\bibitem{note} Here, for specificity, we consider $e$-base logarithms, but the relations are modified only slightly for arbitrary bases.

\bibitem{szedegy-}M. Szegedy, in \href{https://ieeexplore.ieee.org/document/1366222/}{Proceedings of the 45th IEEE Symposium on Foundations of Computer Science (2004), 32}.


\bibitem{q-coherence} A. Streltsov, G. Adesso, and M. B. Plenio, \href{https://doi.org/10.1103/RevModPhys.89.041003}{Rev. Mod. Phys. \textbf{89}, 041003 (2017)}.

\bibitem{brunpra2003} T. A. Brun, H. A. Carteret, and A. Ambainis, \href{https://journals.aps.org/pra/abstract/10.1103/PhysRevA.67.032304}{ Phys. Rev. A \textbf{67}, 032304 (2003)}.

\bibitem {liu} C. Liu and N. Petulante, \href{https://journals.aps.org/pre/abstract/10.1103/PhysRevE.81.031113}{Phys. Rev. E \textbf{81}, 031113 (2010)}.

\bibitem{laflamme2008} C. M. Chandrashekar, R. Srikanth, and R. Laflamme, \href{https://journals.aps.org/pra/abstract/10.1103/PhysRevA.77.032326}{ Phys. Rev. A \textbf{77}, 032326 (2008)}.


\end{thebibliography}

\begin{thebibliography}{99}

\bibitem{book:Arfken} G. Arfken, \emph{Mathematical Methods for Physicists} (Academic Press, San Diego, 1985).

\end{thebibliography}
\end{document}